\documentclass[conference]{IEEEtran}
\IEEEoverridecommandlockouts
\usepackage{cite}
\usepackage{amsmath,amssymb,amsfonts}
\usepackage{algorithmic}
\usepackage{graphicx}
\usepackage{textcomp}
\usepackage{xcolor}
\usepackage{array}
\usepackage{algorithm2e}
\usepackage{floatrow}
\usepackage{caption}
\usepackage{amsthm}
\usepackage{geometry}
\usepackage{subcaption}
\usepackage{setspace}
\usepackage[absolute]{textpos}
\def\r{\mathrm{r}}
\def\u{\mathrm{u}}
\def\s{\mathrm{s}}
\def\b{\mathrm{b}}
\def\c{\mathrm{c}}
\def\x{\mathrm{x}}
\def\y{\mathrm{y}}

\def\h{\mathrm{h}}

\def\t{\mathrm{t}}
\def\B{\mathrm{B}}
\def\U{\mathrm{U}}
\def\L{\mathrm{L}}
\def\R{\mathrm{R}}
\def\I{\mathrm{I}}
\def\H{\mathrm{H}}
\def\G{\mathrm{G}}
\def\A{\mathrm{A}}
\def\M{\mathrm{M}}
\def\P{\mathrm{P}}

\def\BibTeX{{\rm B\kern-.05em{\sc i\kern-.025em b}\kern-.08em
    T\kern-.1667em\lower.7ex\hbox{E}\kern-.125emX}}
\floatstyle{plaintop} 
\restylefloat{table}
\begin{document}

\begin{textblock}{13}(1.5,0.35)
	\noindent X. Liu, L. Xue, S. Sun, and M. Tao, ``Optimization of RIS placement for
satellite-to-ground coverage enhancement," in Proceedings of \textit{2023 IEEE Global Communications Conference Workshop}, 2023. 
\end{textblock}

\title{Optimization of RIS Placement for Satellite-to-Ground Coverage Enhancement\\
{\footnotesize  }

}

\author{
\IEEEauthorblockN{Xingchen Liu, Liuxun Xue, Shu Sun, and Meixia Tao}
\IEEEauthorblockA{Department of Electronic Engineering and the Cooperative Medianet Innovation Center (CMIC),\\ Shanghai Jiao Tong University, Shanghai, China\\Corresponding author: Shu Sun (shusun@sjtu.edu.cn)}
}

\maketitle

\begin{abstract}
In satellite-to-ground communication, ensuring reliable and efficient connectivity poses significant challenges. The reconfigurable intelligent surface (RIS) offers a promising solution due to its ability to manipulate wireless propagation environments and thus enhance communication performance. In this paper, we propose a method for optimizing the placement of RISs on building facets to improve satellite-to-ground communication coverage. We model satellite-to-ground communication with RIS assistance, considering the actual positions of buildings and ground users. The theoretical lower bound on the coverage enhancement in satellite-to-ground communication through large-scale RIS deployment is derived. Then a novel optimization framework for RIS placement is formulated, and a parallel genetic algorithm is employed to solve the problem. Simulation results demonstrate the superior performance of the proposed RIS deployment strategy in enhancing satellite communication coverage probability for non-line-of-sight users. The proposed framework can be applied to various architectural distributions, such as rural areas, towns, and cities, by adjusting parameter settings.
\end{abstract}


\let\thefootnote\relax\footnotetext{This work is supported by the NSF of China under Grant 62271310 and Grant 62125108, and by the  Fundamental Research Funds for Central Universities of China.}

\section{Introduction}

Satellite communication, despite its capacity for coverage and accessibility in remote areas, often encounters degraded signal quality due to obstacles blocking the line-of-sight (LoS) path. The reconfigurable intelligent surface (RIS) has been identified as a potential technology that can extend coverage and improve signal quality by creating links for non-line-of-sight (NLoS) users in satellite-to ground communication.

RIS-assisted terrestrial communication systems have been extensively investigated, especially on RIS's ability to establish indirect LoS links and mitigate signal degradation. For instance, Nemati \textit{et al.} demonstrated the effectiveness of RIS in enhancing 5G coverage\cite{1}. The authors proposed a general tractable approach for analyzing the signal-to-interference ratio (SIR) coverage performance in millimeter-wave cellular networks with RISs and used stochastic geometry to study the average SIR behavior over randomly distributed base stations, RISs, and users in a two-dimensional (2D) space. In \cite{2}, the authors considered a downlink RIS-assisted network with only one base station and one user device. Specifically, an algorithm was proposed to obtain the optimal direction and horizontal distance of the RIS. However, \cite{1} and \cite{2} did not consider the impact of buildings in the actual environment on signal transmission. In \cite{3}, the authors utilized stochastic geometry tools to investigate the impact of large-scale RIS deployment on cellular network performance and presented a series of system-level deployment insights. However, this stochastic-based approach cannot directly provide a deployment plan for RIS in the presence of deterministic building distributions.

In satellite-to-ground communication, RIS integration has been a focal point of research to enhance the propagation channel's reconfigurable radio environment. For example, Lin \textit{et al}. proposed a joint beamforming and optimization scheme for RIS-aided satellite-terrestrial networks to minimize overall transmit power while ensuring user rate requirements\cite{4}. Moreover, Chen \textit{et al.} showcased the potential of RIS-assisted satellite communication to enhance quality and coverage in complex environments\cite{5}. Nevertheless, most existing work assumes a predetermined RIS position with primary focus on phase shift control or joint beamforming and phase shift design, and have not considered the large-scale RIS deployment problem to enhance satellite communication coverage.

In this paper we investigate the impact of large-scale RIS deployment in terrestrial buildings on improving NLoS user coverage in satellite-to-ground communication. We model 3D buildings and propose an RIS deployment algorithm to enhance the average NLoS user coverage under different satellite positions. Additionally, building upon prior research, we derive the lower bounds on the achievable coverage improvement. The main contributions of this work are as follows:

\begin{itemize}
  \item First, we utilize tools from stochastic geometry to derive the theoretical lower bound on the coverage enhancement in satellite-to-ground communication through large-scale RIS deployment, where we consider the impact of randomly distributed RISs within a certain region.
  \item Second, we incorporate the actual positions and 3D orientations of buildings in the real world and propose a novel framework for formulating the optimization problem of the placement of RISs by creating a received power matrix of each satellite position.
  \item Furthermore, we employ the parallel genetic algorithm (PGA) to solve the RIS placement problem and conduct a series of simulation experiments to validate the effectiveness of the PGA for RIS deployment. Simulation results reveal the impact of various factors on the effectiveness of the proposed scheme. Based on this, we provide insights into the RIS deployment strategy.
\end{itemize}

\begin{table}
    \centering
    \footnotesize
    {
    \begin{tabular}{c|c}
    \hline\hline
    \textbf{Notation}     &  \textbf{Description} \\
    \hline
    $\theta_{\min}$     & The minimum satellite elevation angle  \\
    \hline
    $d_{\u,\r}$     & The distance between the user and the RIS  \\
    \hline
    $d_{\r,\s}$     & The distance between the RIS and the satellite  \\
    \hline
    $\lambda_{\b}$     & The PPP density of blockage  \\
    \hline
    $\eta$      & The height distribution difference coefficient  \\
    \hline
    $N_{\B}$    & The number of generated buildings \\
    \hline
    $N_{\U}$  &  The number of users outdoors \\
    \hline
    $K$  & The number of satellite positions traversed\\
    \hline
    $H_{\s}$  & The height of satellite\\
    \hline
    $\epsilon$  & The receive power threshold to be covered\\
    \hline
    $\gamma$     & The RIS deployment ratio\\
    \hline
    \end{tabular}}
    \captionsetup{font=footnotesize}
    \caption{Table of notations}
    \label{tab:my_label}
\end{table}

Most relevant notations are presented in Table I.

\section{System Model}

As shown in Fig. 1, we investigate an RIS-assisted satellite wireless system, where a low-Earth-orbit (LEO) satellite serves multiple users on the ground. The direct LoS links between the satellite and users may be blocked by buildings on the ground; thus RISs are deployed to establish reflecting links. The RISs are placed at the top of the side surfaces of the buildings, facilitating NLoS users in establishing indirect LoS links with satellites, thus enhancing satellite communication coverage rate.

We assume that the LEO satellite communicating with ground users is at the height of $H_\s$. The satellite may appear anywhere from a dome-shaped region with a minimum elevation angle of $\theta_{\min}$ relative to the ground area's center. As the locations of buildings are determined, the NLoS users may change due to the change of satellite position as well as the change of user locations. We assume that users with direct LoS links to the satellite are always coverd, and that an NLoS user is covered as long as its received signal power exceeds a predefined value $\epsilon$. Our goal is to find an RIS deployment strategy to maximize the coverage rate for NLoS users. If no RIS is deployed, the coverage rate for NLoS users is 0.

In our system model, the distribution of the center points of the buildings is assumed to follow a 2D PPP with a Poisson intensity of $\lambda_\b$ within a rectangular region of area $S_\A$. The building's length, width, and height follow three independent uniform distributions within $[L_{\min},L_{\max}]$,$[W_{\min},W_{\max}]$ and $[H_{\min},H_{\max}]$. The mean values of the length and width of the buildings are denoted as $\bar{L}$ and $\bar{W}$, respectively, and their orientation $\omega$ adheres to a uniform distribution in the range of [0, 2$\pi$]. The generated $N_\B$ buildings have four sides, each labeled with a number. 

\begin{figure}[htbp]
\centerline{\includegraphics[width=0.75\linewidth]{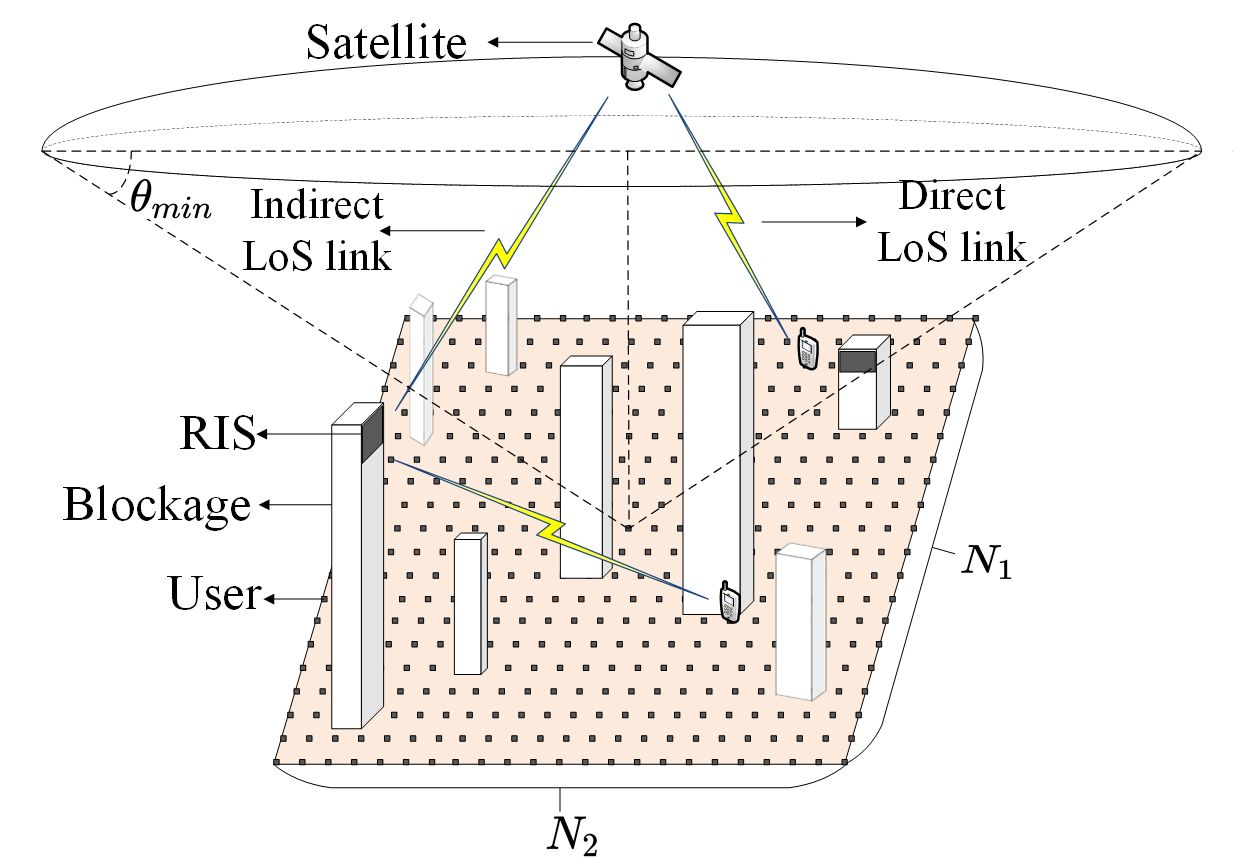}}
\captionsetup{font=footnotesize}
\caption{Satellite communication with the assistance of RISs}
\label{1}
\end{figure}

\section{Analysis of Coverage Probability}

Before presenting our specific algorithm for RIS deployment, we derive the theoretical lower bound on coverage rate of NLoS users for large-scale RIS deployment. Considering a satellite-ground communication network where a typical user is located at the origin with Cartesian coordinates (0, 0, 0). There is a satellite located at coordinates $(x_{\s}, y_{\s}, H_{\s})$ without a direct LoS link to the user (see Fig. 2(a)). We assume that the center points of buildings follow a Poisson point process (PPP) with density $\lambda_{\b}$. According to \cite{6}, if there exists an RIS at coordinates $(x_\r,y_\r,H_\r)$ that can provide an indirect LoS link between the user and the satellite located in the far field of the RIS, then the path loss of the satellite-RIS-user link is proportional to $(d_{\u,\r}d_{\r,\s})^2$, where $d_{\u,\r}$ represents the distance between the user and the RIS, and $d_{\r,\s}$ represents the distance between the RIS and the satellite, i.e.:

\begin{equation}
    \text{PL}=C(d_{\u,\r}d_{\r,\s})^2,
\end{equation}

\noindent where $d_{\u,\r}=\sqrt{x_\r^2+y_\r^2+H_\r^2}$, $d_{\r,\s}=\sqrt{(x_\s-x_\r)^2+(y_\s-y_\r)^2+(H_\s-H_\r)^2}\approx\sqrt{x_\s^2+y_\s^2+(H_\s-H_\r)^2}$. In our model, $C$ is a constant, which can be calculated as \cite{6}:

\begin{equation}
    C=\frac{G_\t G_\r G M^2 N^2 d_\x d_\y c^2 A^2}{64 f_\c^2 \pi^3},
\end{equation}

\noindent where $G_\t$, $G_\r$, and $G$ are the gains of the transmit antenna, receive antenna, and RIS unit cell, respectively, $M$ and $N$ denote the number of rows and columns of unit cells of the RIS, with the size of $d_\x$ and $d_\y$ along two axes, $c$ is the light speed, $f_\c$ is the carrier frequency, and $A$ is the reflection coefficient of the RIS. For simplicity, the influence of the signal's incident and exit angles relative to the RIS on the path loss is omitted.

The blockages are modeled using the line Boolean model \cite{7}, with length $L$, height $H$, and orientation $\omega$. The locations of the midpoints of blockages are modeled as a PPP $\Psi_{\text{bl}}$ with density $\lambda_{\text{bl}}$. Here, we assume that $L$ and $H$ are randomly and uniformly distributed within the range $[L_{\min},L_{\max}]$ and $[H_{\min},H_{\max}]$, respectively, and with probability density distributions and mean values of $f_{\L}(l)$, $\bar{L}$ and $f_{H}(h)$, $\bar{H}$, respectively. The value of $\theta$ represents the angle between the blockage and the positive direction of the $x$-axis, and is assumed uniformly distributed between 0 and 2$\pi$.

For a typical user, due to the high path loss for large propagation distances, we only consider the reflected signals from RISs in the nearby region with a radius of $R$ kilometers, referred to as region $S$. Firstly, we assume that there is only one RIS within region $S$, with an angle $\theta$ between it and the positive x-axis. As shown in Fig. 2b, it can be observed that only when the RIS is deployed in either region $S_1$ or $S_3$, it has the potential to provide an indirect LoS link. Additionally, the probability of the user and the satellite facing the deployment side of the RIS is $\frac{1}{2}$. However, within region $S_2$, the user and the satellite are distributed on the two sides of the RIS, preventing an indirect LoS link.

\begin{figure}
  \centering
  \begin{subfigure}{0.6\textwidth}
    \centering
    \includegraphics[width=\textwidth]{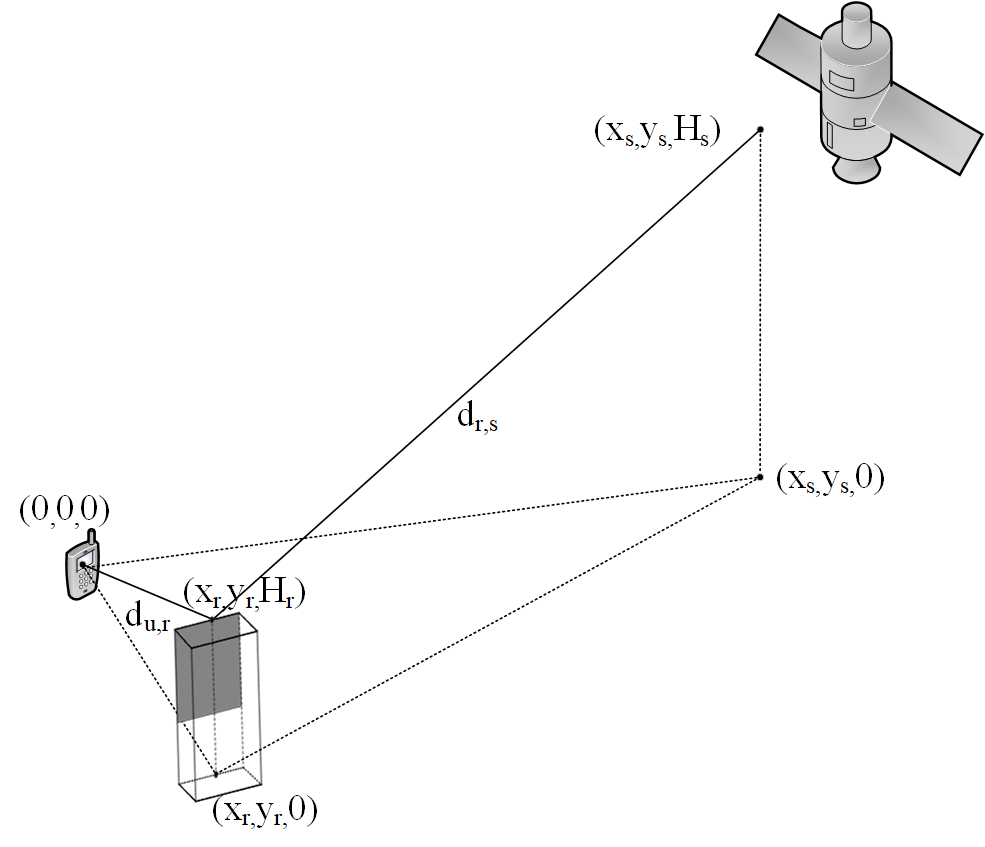}
    \captionsetup{font=footnotesize}
    \caption{3D view}
    \label{fig:subfigure_a}
  \end{subfigure}
  \hfill
  \begin{subfigure}{0.7\textwidth}
    \centering
    \includegraphics[width=\textwidth]{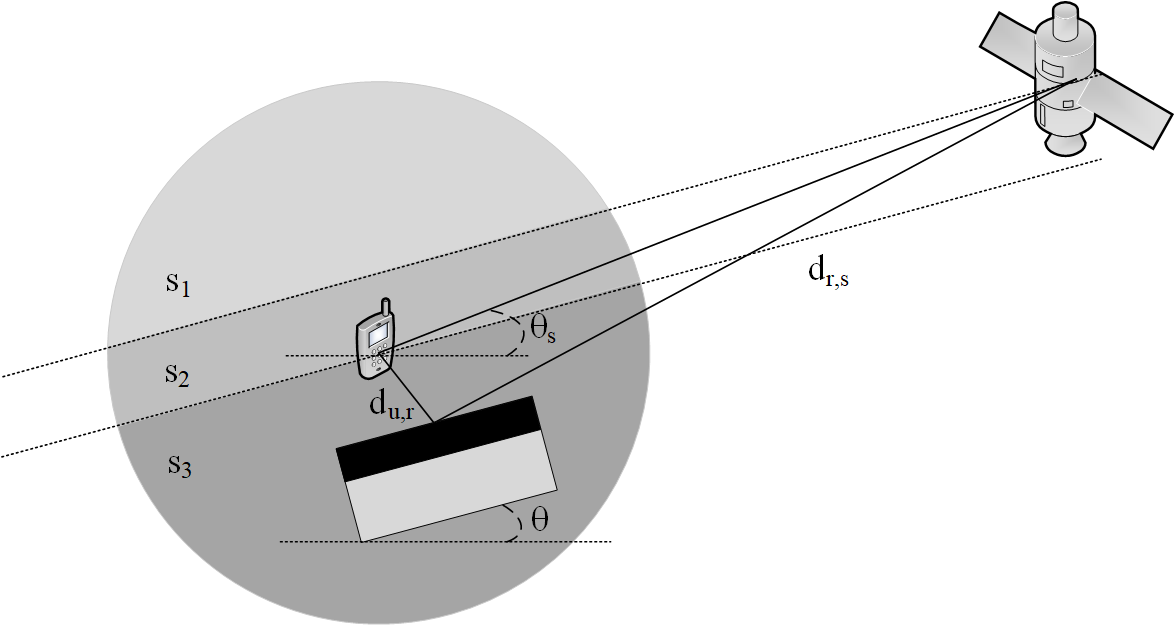}
    \captionsetup{font=footnotesize}
    \caption{Top view}
    \label{fig:subfigure_b}
  \end{subfigure}
  \captionsetup{font=footnotesize}
  \caption{Illustration of the indirect LoS link between the typical user and the satellite.}
  \label{fig:example}
\end{figure}

Hence, the received signal power of the typical user via a randomly distributed RIS with angle $\theta$ can be seen as a random variable:

\begin{equation}
    P_{\r,\theta}=\delta_{1}\delta_{2}\frac{P_{\t}}{C(d_{\u,\r}d_{\r,\s})^2}P_{\text{LoS}}(d_{\u,\r,\x\y})P_{\text{LoS}}(d_{\r,\s,\x\y}),
\end{equation} 

\noindent  where $d_{\u,\r,\rm{xy}}$ and $d_{\r,\s,\rm{xy}}$ are the projection distances from the RIS to the user and from the RIS to the satellite on the XY plane, $P_{\text{LoS}}(r)=\eta \text{exp}(-(\beta r+p))$ is the LoS probability between two points at a distance $r$ \cite{7}, $\beta= \frac{2\lambda_{\text{bl}}(\bar{W}+\bar{L})}{\pi}$, $p=\lambda_{\text{bl}}\bar{W}\bar{L}$ \footnote{When considering the process of forming indirect LoS links with respect to the RIS, we model the buildings as a line Boolean model, without considering the width of the buildings. The buildings are assumed to have only two faces, and a building can deploy at most one RIS. If an RIS is deployed along the longer side, the width of the building's shorter side does not affect the formation of indirect LoS links, and vice versa. However, the width of the building has an impact on the LoS probability between two points. Therefore, when calculating the LoS probability, the influence of the building's width should be considered.}, $\eta$ is a coefficient influenced by the height difference between the transmitter and receiver, $\delta_{1}$ and $\delta_{2}$ are two independent binary random variables that indicate whether the RIS is located in the region $S_1$ or $S_3$, and whether the user and satellite both face the RIS, respectively. If the conditions are met, they take a value of 1. 

In satellite communications, the distance between the satellite and the RIS is much larger than that between the user and the RIS. Therefore, we can assume that $d_{\u,\r}$ is much smaller than $d_{\r,\s}$, and the variation of $d_{\u,\r}$ does not affect the magnitude of $d_{\r,\s}$. In other words, $d_{\r,\s}$ and $d_{\r,\s,\x\y}$ can be considered as a constant. We can regard $d_{\u,\r}$ as a combination of two random variables as $d_{\u,\r}=\sqrt{d_{\u,\r,\x\y}^2+d_{\h}^2}$, and $d_\h$ is the height difference between the RIS and the user. The probability density function (PDF) of $d_\h$ equals $f_\H(h)$. The PDF of $d_{\u,\r,\x\y}$ can be calculated using the properties of PPP, which is as follows:

\begin{equation}
    f_{d_{\u,\r,\x\y}}(r)=\left\{
    \begin{array}{ll}
     \frac{2}{R^2}r & \text{if}\ 0\leq r< R \\
     0 & \text{o.w} \\
    \end{array}
    \right.,
\end{equation}

\noindent where $R$ is the radius of region $S$. Thus the PDF of $d_{\u,\r}$ can be calculated as $f_{d_{\u,\r}}(x)$ using the property of variable substitution for PDFs of random variables.

Next, we derive the PDFs of the random variables  $\delta_{1}$ and  $\delta_{2}$. In satellite communications, for NLoS users, the projected distance from the satellite to the user is much larger than the radius of the region $S$, i.e., the satellite is unlikely to appear directly overhead for NLoS users. When $\theta$ is not within the interval $[\theta_{\s}-\theta_{\t}, \theta_{\s}+\theta_{\t}]$, where $\theta_\t=\text{arcsin}\frac{R}{\sqrt{x_\s^2+y_\s^2}}$ is a very small value, the region $S_1$ disappears, leaving only two equal-sized regions, $S_2$ and $S_3$, each occupying half of the 
region $S$. So we calculate the received power via an RIS in the region $S$ with arbitrary $\theta$, which can be denoted as $P_\r$:

\begin{equation}
    \begin{aligned}
    P_\r &=\frac{1}{2\pi}\int_{0}^{2\pi}P_{\r,\theta}\text{d}\theta \\
        &= \frac{1}{2\pi}\left(\int_{0}^{\theta_{\s}-\theta_{\t}}P_{\r,\theta}\text{d}\theta+\int_{\theta_{\s}-\theta_{\t}}^{\theta_{\s}+\theta_{\t}}P_{\r,\theta}\text{d}\theta+\int_{\theta_{\s}+\theta_{\t}}^{2\pi}P_{\r,\theta}\text{d}\theta\right)\\
        &\overset{\text{(a)}}{\approx}\frac{1}{2\pi-2\theta_\t}\left(\int_{0}^{\theta_{\s}-\theta_{\t}}P_{\r,\theta}\text{d}\theta+\int_{\theta_{\s}+\theta_{\t}}^{2\pi}P_{\r,\theta}\text{d}\theta\right)\\
        & \\
    \end{aligned},
\end{equation}

\noindent where (a) comes from the fact that $\theta_\t$ is very close to 0, therefore the integration over the interval $[\theta_{\s}-\theta_{\t}, \theta_{\s}+\theta_{\t}]$ can be omitted. The PDFs of $\delta_1$ and $\delta_2$ in such a condition follow the same pattern:

\begin{equation}
    f_{\delta_i}(x)=\left\{
    \begin{array}{ll}
     0.5 & \text{if}\ x=0 \\
     0.5 & \text{if}\ x=1 \\
    \end{array}
    \right.,
\end{equation}

\noindent where $i$ = 1, 2. Actually, in the intervals $[0,\theta_\s-\theta_\t]$ and $[\theta_\s+\theta_\t,2\pi]$, the value of $P_{\r,\theta}$ is independent of $\theta$, since both $S_2$ and $S_3$ are half of region $S$. So we have the average received power $\Bar{P}_\r$ as:

\begin{equation}
    \begin{aligned}
    \Bar{P}_\r &=\text{E}_{\delta_1,\delta_2}[P_\r] \\
        &= \text{E}[\delta_1]\text{E}[\delta_2]\frac{P_{\t}}{C(d_{\u,\r}d_{\r,\s})^2}P_{\text{LoS}}(d_{\u,\r,\x\y})P_{\text{LoS}}(d_{\r,\s,\x\y})\\
        &=\frac{1}{4}\frac{P_{\t}}{C(d_{\u,\r}d_{\r,\s})^2}P_{\text{LoS}}(d_{\u,\r,\x\y})P_{\text{LoS}}(d_{\r,\s,\x\y})\\
        &=B\frac{\text{exp}(-\beta d_{\u,\r,\x\y})}{d_{\u,\r}^2},
    \end{aligned}
\end{equation}

\noindent where $B=\eta_{1}\eta_{2}\frac{P_{\t}\text{exp}(-\beta d_{\r,\s,\x\y}-2p)}{4Cd_{\r,\s}^2}$ is a constant, $\eta_{1}$ and $\eta_{2}$ are height difference factors between the satellite and RIS, and between the RIS and users, respectively. The PDF of $\Bar{P}_\r$ can also be calculated using the property of variable substitution for PDFs of random variables as $f_{\Bar{P}_\r}(x)$.

Up to now, we have derived the PDF of the average received power when a single RIS is deployed in the region $S$. If there are $N_\R$ RISs deployed in the region $S$, the PDF of the average received power can be calculated as:

\begin{equation}
    f_{\Bar{P}_\r,N_\R}(x)=\underbrace{f_{\Bar{P}_\r}(x) \ast f_{\Bar{P}_\r}(x) \ast \cdot \cdot \cdot \ast f_{\Bar{P}_\r}(x)}_{\text{$N_\R$}},
\end{equation}

\noindent where '*' represents the convolution operator. As a result, the probability that the average received power via $N_\R$ RISs in the region $S$ is larger than a predefined threshold $\epsilon$ is:

\begin{equation}
    P_{\text{cov}}=\int_{\epsilon}^{\infty}f_{\Bar{P}_\r,N_\R}(x)\text{d}x,
\end{equation}

\noindent which can be numerically computed when system parameters are given. The above derivation assumes that the RISs are uniformly distributed in the nearby region without any specific deployment strategy; hence (9) provides a lower bound on the coverage probability of NLoS users. The following section will introduce an optimization problem to maximize the coverage rate in a practical scenario, where the positions of RISs are well-designed.

\section{RIS Deployment Strategy}

We consider the case where $N_1 \times N_2$ users are uniformly distributed in the aforementioned rectangular region. Users inside buildings are removed, resulting in $N_\U = N_1 \times N_2 - N_\I$ outdoor users, where $N_I$ represents the number of users inside buildings. This finite set of discrete points can emulate the actual user distribution with a large density.

 As an LEO satellite can appear anywhere within the predefined dome-shape region, we generate $K$ satellite positions using the Fibonacci grid method and calculate the received power matrix $\mathbf{W}_l^k$ for each user, which will be detailed in the sequel. These matrices under different satellite positions are combined for joint optimization, in order to improve the robustness of optimization results concerning the possible presence of satellites at arbitrary positions in real scenarios. Under the $k$-th satellite position, there exist $L^k$ NLoS users for optimization.

To represent the impact of the RISs at different locations on the $l$-th NLoS user under the $k$-th satellite position, we create a matrix $\mathbf{W}_{l}^{k}$ with dimensions of $N_\B \times 4$. The element in the $i$-th row and $j$-th column, $\mathbf{W}_{l_{i,j}}^{k}$, represents the received power of the $l$-th user through the cascaded channel formed by the RIS when it is deployed on the $j$-th surface of the $i$-th building, calculated by $\mathbf{W}_{l_{i,j}}^{k}=\frac{P_\t}{C(d_{\u,\r}d_{\r,\s})^2}$. However, if there is no LoS link between the RIS and $l$-th NLoS user or between the RIS and the satellite, or $d_1>R$, we have $\mathbf{W}_{l_{i,j}}^{k}=0$.

In the optimization problem, our optimization variable $\mathbf{X}$ is a matrix of size $N_\B \times 4$. Each entry $\mathbf{X}_{ij}$ is a binary variable that takes the value 1 when the RIS is deployed on the $j$-th surface of the $i$-th building and 0 otherwise. The optimization problem is formulated as follows:

\begin{equation}
\begin{aligned}
& \underset{\mathbf{X}}{\text{max}} & & \frac{\sum_{k=1}^{K} \sum_{l=1}^{L^k} u((\sum_{i=1}^{N_\B} \sum_{j=1}^4 (\mathbf{W}_{l}^{k} \cdot \mathbf{X})_{ij}-\epsilon))}{\sum_{k=1}^{K} L^k},  \\
& \text{s.t.} & & \sum_{i=1}^{N_\B} \sum_{j=1}^4 \mathbf{X}_{ij} \leq N_B\gamma \\
& & & \textbf{X}[1\ 1\ 1\ 1]^T \leq ([1\ 1\ ...\ 1]_{1\times N_\B})^T,
\end{aligned}
\end{equation}

\noindent where $u(x)$ is the step function. $\sum_{i=1}^{N_B} \sum_{j=1}^4 (\mathbf{W}_{l}^{k} \cdot \mathbf{X})_{ij}$ is the received power of the $l$-th NLoS user under the $k$-th satellite position via all RISs within radius $R$. We subtract $\epsilon$ from the received power of the user and then apply the step function. The parameter $\gamma$ represents the RIS deployment ratio between 0 and 1. The numerator of the objective function represents the number of NLoS users covered, while the denominator represents the total number of NLoS users (including all satellite positions traversed). The objective of the optimization problem is to find the optimal RIS deployment matrix $\mathbf{X}$ that maximizes the average coverage probability for NLoS users across various satellite positions, subject to the constraint that the total number of deployed RISs does not exceed $N_\B \times \gamma$ and each building deploys at most one RIS.

  \begin{algorithm}
    \setstretch{0.85}
    \SetAlgoLined
    \SetKwInOut{Input}{Input}
    \SetKwInOut{Output}{Output}
    \hrule
    \BlankLine
    Algorithm 1: Parallel Genetic Algorithm
    \BlankLine
    \hrule
    \BlankLine
    \Input{Number of populations $N_\P$, population size $S_\P$, number of generations $N_\G$, migration interval $I$, migration times $N_\M$, elite selection size $E_1$ and $E_2$}
    \Output{Optimal solution for the given problem}
  
    \BlankLine
    Initialize $N_\P$ populations randomly\;
    \For{$m=1$ to $N_\M$}{
      \For{$i=1$ to $I$}{
        Conduct genetic algorithm on each population\;
      }
      \For{$p=1$ to $N_\P$}{
        $\cdot$ Evaluate fitness for individuals in population $p$\;
        $\cdot$ Sort population $p$ based on fitness\;
        $\cdot$ Select top $E_1$ individuals as parents for crossover in each neighboring population\;
        $\cdot$ Perform crossover and mutation on parents to generate offspring\;
        $\cdot$ Replace the worst $E_1$ individuals with the offspring\;
      }
      \For{$p=1$ to $N_\P$}{
        $\cdot$ Select the best $E_2$ individual for mutation replacement\;
        $\cdot$ Perform mutation on the selected individual\;
        $\cdot$ Replace the worst individual with the mutated individual in the next population\;
      }
    }
    \hrule
  \end{algorithm}

Given the non-linearity of our optimization problem, traditional optimization methods may struggle to find an optimal solution. Genetic algorithm (GA) is a population-based search algorithm that can effectively handle complex and non-linear optimization problems by exploring a diverse solution space. Moreover, the parallel genetic algorithm (PGA), a modified variant of GA, provides the advantage of parallel processing, which allows for faster convergence and better exploration of the solution space\cite{8}. Therefore, we select PGA as our optimization algorithm to leverage its superior ability to handle binary decision variables, such as our $\mathbf{X}$ matrix. In our application, the PGA leverages the concept of multiple populations, each evolving independently yet occasionally exchanging information, which is shown in Algorithm 1. 

After obtaining the optimized deployment scheme for RISs through satellite positions traversed, we will proceed to test its performance. The specific approach involves randomly generating multiple sets of satellite positions and calculating their average NLoS user coverage to verify the stability and robustness of the previously obtained RIS deployment scheme. In this process, the satellite positions obtained from the Fibonacci grid method and used to solve the optimization problem can be considered as the training set, while multiple sets of randomly generated satellite positions can be considered as the testing set.

\section{Numerical Results}

In this section, we evaluate the performance of our proposed RIS deployment strategy via numerical simulations.

We consider a 1 km $\times$ 1 km urban area with building parameters set as follows: the primitive Poisson density of center points of buildings is $\lambda_{\b}^{'}$, building height ranging from 80 m to 120 m, and building length and width ranging from 30 m to 40 m. After removing the overlapping buildings, there are $N_{\B}$ remaining buildings. We generate 30 $\times$ 30 users on the XY plane and remove the users inside the buildings, resulting in $N_{\U}$ users (approximately around 800). Users can be influenced by RISs within $R=0.5$ km. The satellite's height $H_{\s}$ is set to 600 km. Satellite positions are generated in a dome-shaped region with a minimum elevation angle of $\theta_{\min}=30^{\circ}$ relative to the ground area's center, and the number of satellite positions traversed is $K$. 

\begin{figure}[htbp]
\centerline{\includegraphics[width=0.9\linewidth]{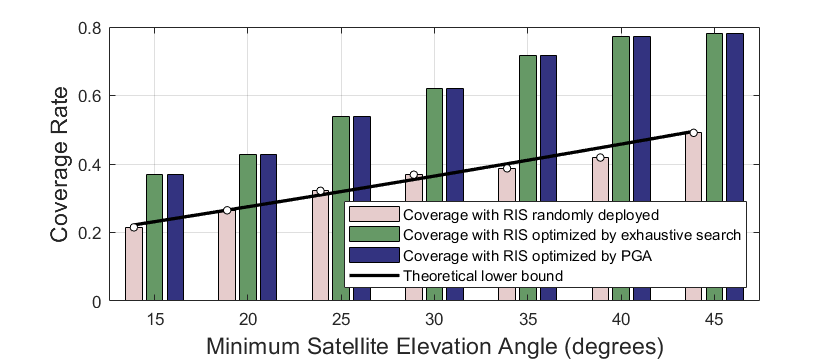}}
\captionsetup{font=footnotesize}
\caption{Comparison of the coverage rate for NLoS users between PGA and exhaustive search.}
\label{Comparison}
\end{figure}

\begin{figure}[htbp]
\centerline{\includegraphics[width=0.9\linewidth]{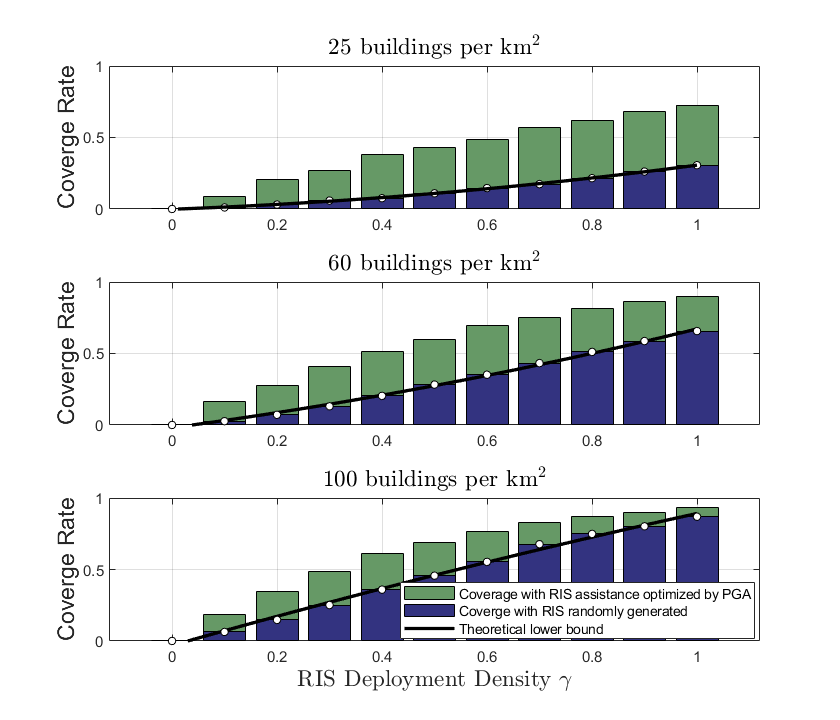}}
\captionsetup{font=footnotesize}
\caption{Coverage rate for NLoS users under different RIS deployment ratios and building environments.}
\label{BE}
\end{figure}

\subsection{Simulation Setup}

To compare with the lower bound of the coverage probability in (9), we need to calculate the equivalent parameters. Since there are $N_\B$ remaining buildings after removing the overlapping buildings in the simulation experiments, we can define $\lambda_\b$ as $\frac{\Bar{N}_\B}{\pi R^2}$, where $\Bar{N}_\B$ represents the average number of blockages within $R$  km for each NLoS user. The threshold $\epsilon$ is set to $10^{-3}$W, which is determined by the expected maximum bit error rate.

\subsection{Results and Discussion}
In this section, we investigate the impact of several factors on the system performance, including minimum satellite elevation angle, RIS deployment ratio, building density, and number of satellite positions traversed. We present the simulation results to demonstrate these effects and analyze their implications on our proposed RIS deployment strategy.

\begin{figure}[htbp]
\centerline{\includegraphics[width=0.9\linewidth]{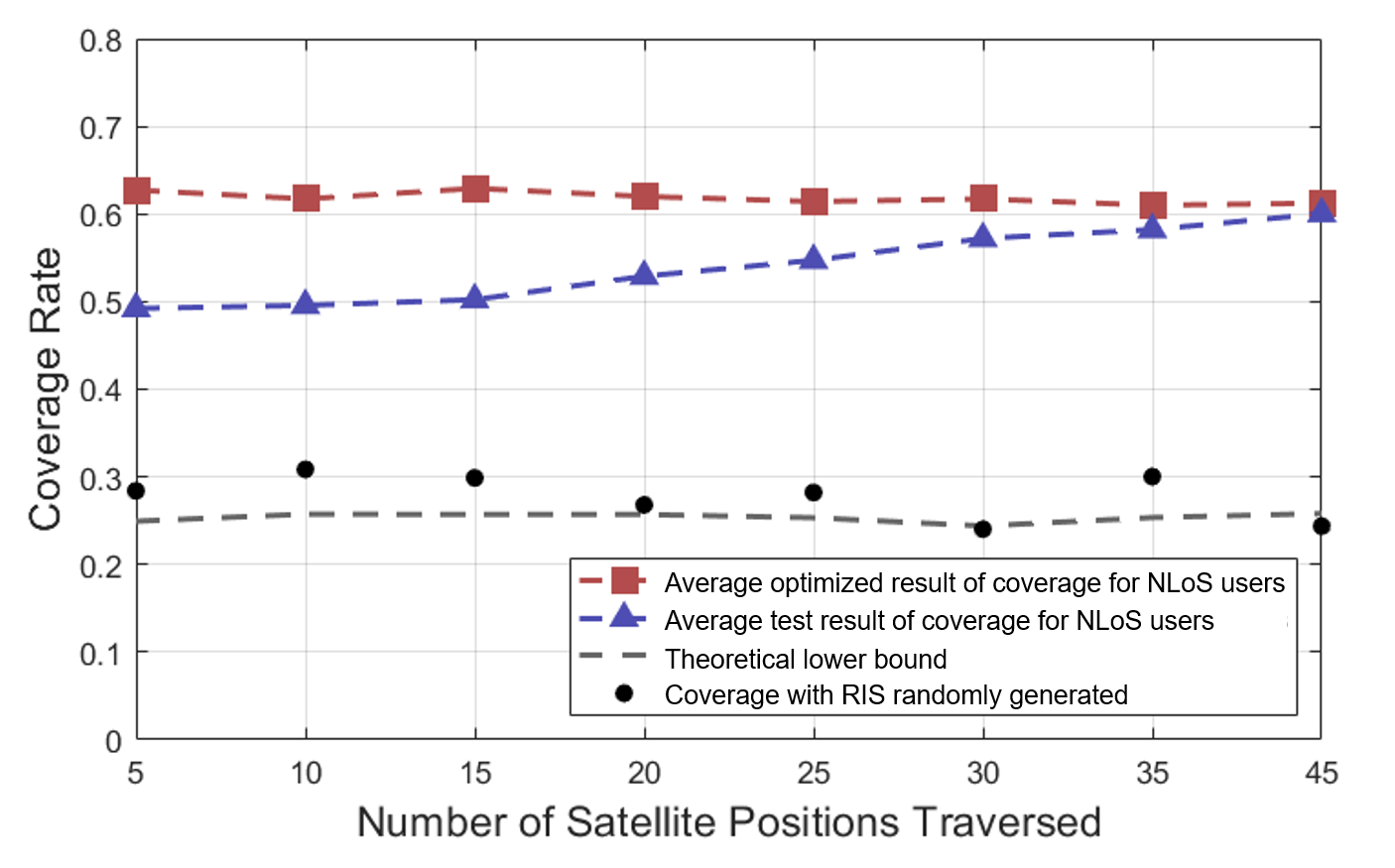}}
\captionsetup{font=footnotesize}
\caption{Training and testing results under different numbers of satellite positions traversed.}
\label{TT}
\end{figure}

\subsubsection{Comparison with Exhaustive Search}
We validate the effectiveness of the PGA for RIS deployment by comparing it with the exhaustive search method. The RIS deployment ratio $\gamma$ is set to be 0.5, and the number of satellite positions traversed $K$ is set to be one\footnote{The satellite is positioned precisely at the edge of the dome-shaped area, with an elevation angle $\theta_{min}$ relative to the center of the ground area, facilitating our investigation of the impact of different elevation angles on deployment efficiency.}. Experiments were conducted on the two methods for a relatively small area involving eight buildings. As shown in Fig. \ref{Comparison}, the results demonstrate that the PGA achieved a result comparable to the exhaustive search. The running time of exhaustive search is thousands of times, or even more, than that of the PGA. This result highlights the reliability and efficiency of the PGA in solving such problems. Moreover, as the elevation angle decreases, the enhancement in NLoS users coverage due to RIS deployment becomes smaller. This is because, with smaller elevation angles, buildings impose stronger obstruction on users. Additionally,  the theoretical lower bound, equivalent to randomly deploying RISs on buildings, fits well with the simulation, which demonstrates the applicability of our theoretical lower bound to simulate real-world scenarios. The optimized large-scale RIS deployment scheme significantly outperforms the theoretical lower bound, demonstrating the superiority of our algorithm.

\subsubsection{Impact of RIS Deployment Ratio and Building Density}
In this experiment, the number of satellite positions traversed $K$ is set to 30. From Fig. 4, it can be observed that, regardless of the type of building density, an increase in the RIS deployment ratio significantly improves user coverage. When the building density is 25 buildings per square kilometer, it can be noticed that the increase in coverage due to RIS deployment is relatively small. This is because, with the same RIS deployment ratio, the absolute number of RISs is also small when there are fewer buildings. We also observe that when the coverage is not very high, both the coverage achieved through PGA optimization and the coverage obtained by randomly placing RISs exhibit approximately linear growth with increasing RIS deployment density $\gamma$.

\subsubsection{Impact of the number of satellite positions traversed}
In this simulation experiment, we investigate the impact of varying the number of satellite positions traversed on NLoS user coverage under testing satellite positions. Here, $\gamma$ is fixed at 0.5. The results in Fig. 5 revealed a positive correlation between the number of satellite positions traversed and the NLoS user coverage during testing. With more satellite positions traversed, better results can be seen in the test process and the test results become closer to the optimized results. Therefore, we can consider the optimized results as an approximate upper bound of the test results. These findings emphasize the importance of considering an appropriate number of satellite positions traversed, which can achieve a balance between improving algorithm accuracy and saving computational costs.

\section{Conclusion}

In this study, we proposed an RIS deployment strategy to enhance coverage for NLoS users in LEO satellite communication systems. We derived the theoretical lower bound of NLoS user coverage probability and formulated a non-linear binary optimization problem to achieve optimal RIS deployment. The complex nature of the problem led us to employ the PGA, which proved efficient and reliable compared to exhaustive search methods. Numerical simulations have revealed that an increased RIS deployment ratio can significantly improve user coverage, and both the theoretical lower bound and the optimized results exhibit approximate linear increase within a certain range as $\gamma$ increases. Additionally, more satellite positions traversed result in better coverage performance, with an approximate upper bound based on optimized results. Overall, the proposed RIS deployment strategy offers an effective solution for improving NLoS user coverage in LEO satellite networks. Future studies can focus on exploring cooperative schemes, RIS-assisted MIMO transmissions, and the integration of machine learning in RIS placement optimization.

\end{document}